\documentclass[a4paper,11pt]{article}
\usepackage{pos}

\usepackage{lineno}
\usepackage[colorinlistoftodos]{todonotes}

\title{Indirect search for dark matter in the Galactic Centre with IceCube}
\ShortTitle{IceCube indirect GC DM search}

\author{The IceCube Collaboration \\{\normalsize \normalfont(a complete list of authors can be found at the end of the proceedings)}}

\emailAdd{nadege.iovine@icecube.wisc.edu}

\abstract{
Even though there are strong astrophysical and cosmological indications to support the existence of dark matter, its exact nature remains unknown. We expect dark matter to produce standard model particles when annihilating or decaying, assuming that it is composed of Weakly Interacting Massive Particles (WIMPs). These standard model particles could in turn yield neutrinos that can be detected by the IceCube neutrino telescope. The Milky Way is expected to be permeated by a dark matter halo with an increased density towards its centre. This halo is expected to yield the strongest dark matter annihilation signal at Earth coming from any celestial object, making it an ideal target for indirect searches. In this contribution, we present the sensitivities of an indirect search for dark matter in the Galactic Centre using IceCube data. This low energy dark matter search allows us to cover dark matter masses ranging from 5~GeV to 1~TeV. The sensitivities obtained for this analysis show considerable improvements over previous IceCube results in the considered energy range.\\

\vspace{4mm}
{\bfseries Corresponding authors:}
Nadège Iovine$^{1*}$, Juan A. Aguilar$^{1}$\\
{$^{1}$ \itshape Université Libre de Bruxelles}\\
$^*$ Presenter

\FullConference{37$^{\rm{th}}$ International Cosmic Ray Conference (ICRC 2021)\\
		July 12th -- 23rd, 2021\\
		Online -- Berlin, Germany}
}

\begin{document}
\maketitle


\section{Introduction}\label{sec:intro}
The existence of dark matter is well-supported by a variety of astrophysical and cosmological indications~\cite{Bertone:2016nfn,Ade:2013zuv}. Observations suggest that galaxies are surrounded by a halo of thermal relic dark matter whose density increases towards the centre of the galaxy~\cite{Kravtsov:1997dp}. Under the assumption that dark matter consists of Weakly Interacting Massive Particles (WIMPs)~\cite{Steigman:1984ac}, this high concentration of dark matter at the centre of galaxies would favour the annihilation of dark matter particles. When annihilating or decaying, we expect these dark matter particles to produce standard model particles, which could in turn yield stable particles such as neutrinos. Indirect dark matter experiments, like neutrino telescopes, aim to detect these resulting particles. The presented analysis searches for neutrinos from dark matter self-annihilation in the Galactic Centre using the IceCube~\cite{Aartsen:2016nxy} detector. With this analysis, the aim is to improve the detection potential for dark matter masses ranging from 5~GeV to 1~TeV.

\section{Dark Matter Phenomenology}
\label{sec:DM}

\begin{figure}[!tbp]
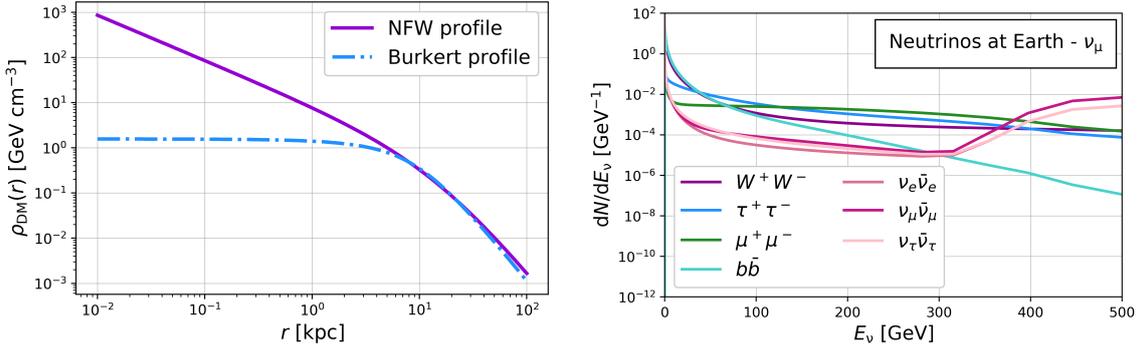

    \centering
    \begin{minipage}[b]{.49\textwidth}
    \includegraphics[width=\linewidth]{Figures/DM_density_Burkert_NFW.pdf}
    \end{minipage}
    \hfill
    \begin{minipage}[b]{.49\textwidth}
    \includegraphics[width=\linewidth]{Figures/PPPC4Spectra_atEarth_500GeV.pdf}
    \end{minipage}
    \caption{\textbf{Left:} Dark matter density, $\rho_{\mathrm{DM}}$, as a function of the distance to the Galactic Centre, $r$, for the NFW and Burkert profiles. \textbf{Right:} Differential number of muon neutrinos per annihilation at Earth for a dark matter mass of 500~GeV.}
    \label{fig:Model_parameters}
\end{figure}

We can deduce the expected differential neutrino flux from dark matter self-annihilation in the Galactic Centre from the following equation~\cite{Yuksel:2007ac}:

\begin{equation}
    \frac{\mathrm{d}\phi_{\nu}}{\mathrm{d}E_{_{\nu}}} = \frac{1}{4\pi} \, \frac{\langle \sigma_{\mathrm{A}} \upsilon \rangle}{2 \, m_{\mathrm{DM}}^2} \; \frac{\mathrm{d}N_{\nu}}{\mathrm{d}E_{\nu}} \; J_{\Psi} \, ,
    \label{eq:sig_expectation}
\end{equation}

\noindent where $m_{\mathrm{DM}}$ is the dark matter mass and $\langle \sigma_{\mathrm{A}} \upsilon \rangle$ is the thermally-averaged dark matter self-annihilation cross-section. The differential number of neutrinos produced per annihilating pair of dark matter particles, $\mathrm{d}N_{\nu} / \mathrm{d}E_{\nu}$, is taken from the PPPC4 tables~\cite{Cirelli:2010xx} and a 100$\%$ branching ratio into either $\nu_{e}\bar{\nu}_{e}$, $\nu_{\mu}\bar{\nu}_{\mu}$, $\nu_{\tau}\bar{\nu}_{\tau}$, $W^+W^-$, $\mu^+\mu^-$ $\tau^+\tau^-$, or $b\bar{b}$ is assumed. These annihilation channels were selected as the corresponding spectra cover a wide range, with the softest spectrum given by $b\bar{b}$ to the hardest spectra given by the $\nu_{i}\bar{\nu}_{i}$ channels, where $i$ indicates the neutrino flavour.
This analysis also probes dark matter masses ranging from 5~GeV to 1~TeV.
Neutrino oscillations between the source and the Earth is taken into consideration and the spectra at Earth for all annihilation channels considered and a dark matter mass of 500 GeV can be seen in Figure~\ref{fig:Model_parameters}.
The J-factor, $J_{\Psi}$, is defined as the integral over the solid angle, $\Delta \Omega$, and line of sight (l.o.s) of the squared dark matter density, $\rho_{\mathrm{DM}}$:

\begin{equation}
    J_{\Psi} = \int_{\Delta \Omega} \mathrm{d}\Omega(\Psi) \int_{\mathrm{l.o.s}} \rho_{\mathrm{DM}}^2\left(r(l, \Psi)\right) \, \mathrm{d}l \, , 
    \label{eq:J_factor}
\end{equation}

\noindent where $\Psi$ is the opening angle to the Galactic Centre. The distribution of dark matter density in the Milky Way can be expressed as a function of the distance, $r$, to the Galactic Centre following~\cite{An:2012pv}:

\begin{equation}
    \rho_{\mathrm{DM}}(r) = \frac{\rho_0}{\left(\delta + \frac{r}{r_{\mathrm{s}}}\right)^{\gamma} \cdot \left[1+\left(\frac{r}{r_{\mathrm{s}}}\right)^{\alpha}\right]^{(\beta-\gamma)/\alpha}} \; .
    \label{eq:DM_density}
\end{equation}

\noindent Values for the normalisation density, $\rho_0$, and the scale radius, $r_s$, are taken from~\cite{Nesti:2013uwa}. As the choice of halo model strongly influences the resulting J-factor, we consider two different dark matter halo model for this analysis. Both models can be expressed by~\ref{eq:DM_density}, for which the parameters $(\alpha,\beta,\delta,\gamma)$ are equal to (2,3,1,1) for the Burkert~\cite{Burkert:1995yz} profile and (1,3,0,1) for the Navarro-Frenk-White profile (NFW)~\cite{Navarro:1995iw}.
For ``cuspy'' dark matter halo profiles, such as the NFW profile,  the dark matter density distribution peaks significantly towards the centre of the galaxy, making the signal signature easier to distinguish from the background. As a result, the NFW profile leads to more optimistic sensitivities than core dark matter profiles, such as the Burkert profile. 
The corresponding dark matter densities can be seen in Figure~\ref{fig:Model_parameters}. The J-factors as a function of the opening angle to the Galactic Centre, $\Psi$, are computed using \emph{Clumpy}~\cite{Hutten:2018aix} for both halo profiles.

\section{Event selection}
\label{sec:event_selection}
Atmospheric muons and neutrinos created by the interactions of cosmic rays in the upper atmosphere represent the main background of IceCube. For up-going events, the Earth acts as a shield against atmospheric muons, which significantly reduces the background. However, when considering sources above the horizon, such as the Galactic Centre, a veto is required in order to reject atmospheric muons. 
For this analysis, a pre-existing event selection, called oscNext, is considered. This low energy data set is optimised for atmospheric neutrino oscillation measurements. 
For this event selection, only events recorded within the DeepCore sub-detector~\cite{Collaboration:2011ym} are selected, while the remaining parts of the IceCube detector are used as a veto. The oscNext event selection consists of DeepCore events recorded from 2012 to 2020, for a total livetime of 8.03 years.

\section{Analysis Method}
\label{sec:analysis_method}
We consider a binned likelihood method in order to search for an excess of signal neutrinos in the Galactic Centre. This algorithm compares the observed data distribution to expectations based on the background and signal distributions, for each considered combination of dark matter mass, annihilation channel and halo profile. The distributions of interest are used as probability distribution functions (PDFs). 
For this search, we use three-dimensional PDFs in which the angular information is considered alongside information about the energy and the neutrino flavour of the event. Therefore, the three dimensions of the PDFs are the opening angle to the Galactic Centre ($\Psi_{\mathrm{reco}}$), the energy ($\log_{10}(E_{\mathrm{reco}})$) and the reconstructed neutrino flavour, know as the particle ID (PID).
The background PDF is built from Monte Carlo simulations weighted according to the expected atmospheric flux. The signal PDFs are also constructed from generic neutrino simulations which are weighted with the source morphology and the neutrino spectrum according to Equation~\ref{eq:sig_expectation}.
A distinct signal PDF is computed for each of the combination of dark matter mass, annihilation channel and halo profile. For the $\nu_{e}\bar{\nu}_{e}$, $\nu_{\mu}\bar{\nu}_{\mu}$ and $\nu_{\tau}\bar{\nu}_{\tau}$ channels, dark matter masses above 200~GeV are not considered as the signal peaks at $E_{\mathrm{reco}}$ where the contribution of the background is close to zero. In order to avoid such scenario, we introduce a cut restricting the possible parameter combinations. For each combination of dark matter mass, annihilation channel and halo profile, we compute the weighted median of the distribution in reconstructed energy, $E_{\mathrm {reco}}$. If the resulting median is above the upper bound of the region containing 95$\%$ of the background, the corresponding signal combination is discarded. This is the case for masses above 200~GeV for dark matter annihilation through $\nu_{i}\bar{\nu}_{i}$, where $i$ represents the neutrino flavour.

\begin{figure}
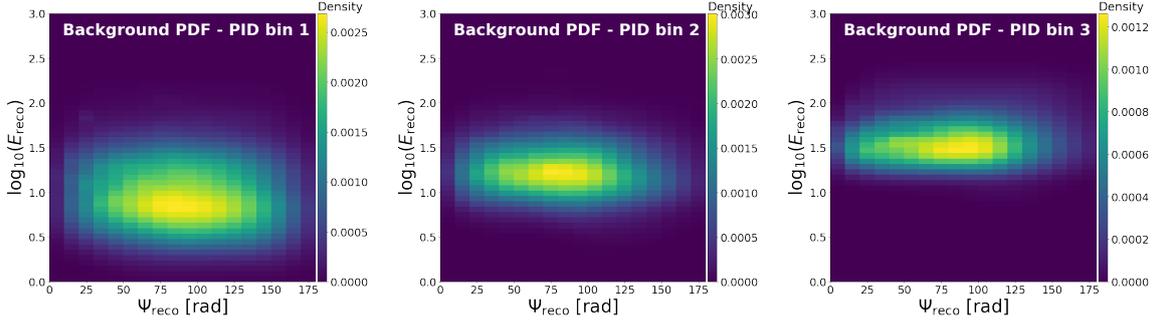

    \centering
    \begin{minipage}[b]{.32\textwidth}
    \includegraphics[width=\linewidth]{Figures/PDF_BG_KDE_scott_PIDbin1.pdf}
    \end{minipage}
    \hfill
    \centering
    \begin{minipage}[b]{.32\textwidth}
\includegraphics[width=\linewidth]{Figures/PDF_BG_KDE_scott_PIDbin2.pdf}
    \end{minipage}
    \hfill
    \centering
    \begin{minipage}[b]{.32\textwidth}
    \includegraphics[width=\linewidth]{Figures/PDF_BG_KDE_scott_PIDbin3.pdf}
    \end{minipage}
    \caption{Two-dimensional projection of the background PDF in $\Psi_{\mathrm{reco}}$ and $\log_{10}(E_{\mathrm{reco}})$ for each of the 3 PID bins described earlier.}
    \label{fig:BG_PDF}
\end{figure}

\begin{figure}
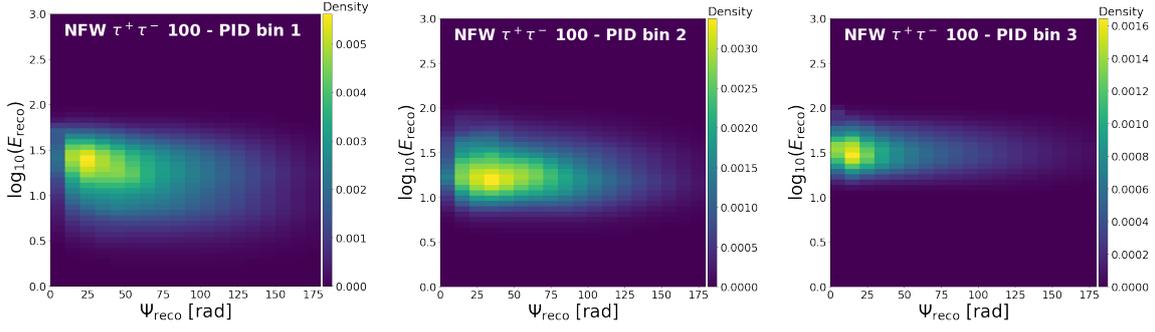

    \centering
    \begin{minipage}[b]{.32\textwidth}
    \includegraphics[width=\linewidth]{Figures/PDF_Signal_KDE_scott_NFW_tautau_100_PIDbin1.pdf}
    \end{minipage}
    \hfill
    \centering
    \begin{minipage}[b]{.32\textwidth}
    \includegraphics[width=\linewidth]{Figures/PDF_Signal_KDE_scott_NFW_tautau_100_PIDbin2.pdf}
    \end{minipage}
    \hfill
    \centering
    \begin{minipage}[b]{.32\textwidth}
    \includegraphics[width=\linewidth]{Figures/PDF_Signal_KDE_scott_NFW_tautau_100_PIDbin3.pdf}
    \end{minipage}
    \caption{Two-dimensional projection of the signal PDF in $\Psi_{\mathrm{reco}}$ and $\log_{10}(E_{\mathrm{reco}})$ for each of the 3 PID bins described earlier, assuming the annihilation of dark matter particles with masses equal to 100~GeV through the $\tau^+\tau^-$ channel for the NFW halo profile.}
    \label{fig:Signal_PDF}
\end{figure}

The binning in PID is chosen to optimise the separation between track and cascade events, resulting in 3 PID bins with edges defined as [0, 0.5, 0.85, 1]. Values of the PID close to zero indicates cascade-like events, while values close to one suggest that the event is track-like. This choice of binning thus gives us a first bin containing cascade-like events, a middle bin consisting of a mixture of tracks and cascades, as well as a third bin with track-like events.
In order to smooth the background and signal distributions, a Kernel Density Estimation (KDE) is used. To estimate our probability density functions, a Gaussian kernel is thus applied on the $\log(\Psi_{\mathrm{reco}})$ and $\log_{10}(E_{\mathrm{reco}})$ distributions, for each PID bin. The resulting smoothed distributions are then binned with 18 bins in $\Psi_{\mathrm{reco}}$ ranging from $0^{\circ}$ to $180^{\circ}$ and 50 bins in $\log_{10}(E_{\mathrm{reco}})$ between 0 and 3. The two-dimensional projections of the background PDF for each PID bin can be seen in Figure~\ref{fig:BG_PDF}. Similarly, the signal PDF for dark matter particles with a mass of 100 GeV annihilating through the $\tau^+\tau^-$ channel is shown in Figure~\ref{fig:Signal_PDF}, assuming the NFW halo profile.

The considered likelihood function is built as the product of the Poisson probabilities to observe $n_{\mathrm{obs}}^i$ events in a specific bin $i$:

\begin{equation}
    \mathcal{L}(\mu) = \prod_{i=min}^{max} \frac{\left(n_{\mathrm{obs}}^{\mathrm{tot}} \, f^{i}(\mu)\right)^{n^{i}_{\mathrm{obs}}}}{n^{i}_{\mathrm{obs}}!} e^{-n_{\mathrm{obs}}^{\mathrm{tot}} \, f^{i}(\mu)}\, ,
    \label{eq:Likelihood}
\end{equation}

\noindent where $n_{\mathrm{obs}}^{\mathrm{tot}}$ is the total number of events in the sample and $\mu \in [0,1]$ is the fraction of signal events of the total sample.
The fraction of events within a bin $i$ is expressed as:

\begin{equation}
    f^{i}(\mu) = \mu \, f_s^i \, + \, (1 - \mu) \, f^i_{\mathrm{bg}} \, ,
    \label{eq:Signal_fraction}
\end{equation}

\noindent where $f_{\mathrm{bg}}$ and $f_s$ are the background and signal PDFs. The best estimate of the signal fraction, $\mu_{\mathrm{best}}$, is obtained by maximising the likelihood, $\mathcal{L}(\mu)$. If this value is consistent with the background-only hypothesis, the upper limit on the signal fraction at the $90\%$ confidence level (CL), $\mu_{\mathrm{90}}$, is computed according to the likelihood interval method~\cite{Cowan:2010js}. The $90\%$ CL sensitivity is computed by generating 100,000 pseudo-experiments sampled from the background-only PDF. We quote as $90\%$ CL sensitivity, $\hat{\mu}_{\mathrm{90}}$, the median value of the upper limits obtained for each of these pseudo-experiments. From this, we can deduce the sensitivity on the thermally-averaged dark matter self-annihilation cross-section, $\langle \sigma_{\mathrm{A}} \upsilon \rangle$, using the total number of observed events and the estimated number of events for a specific combination of dark matter mass, annihilation channel and halo profile.

\section{Sensitivities}
\label{sec:sensitivities}
\begin{figure}
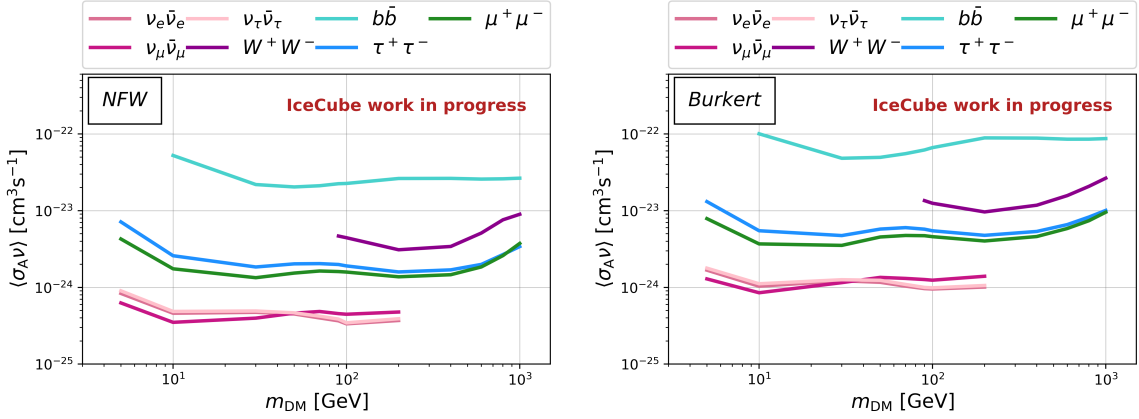

    \centering
    \begin{minipage}[b]{.49\textwidth}
    \includegraphics[width=\linewidth]{Figures/Sensitivities_AllChannels_KDE_scott_NFW.pdf}
    \end{minipage}
    \hfill
    \begin{minipage}[b]{.49\textwidth}
    \includegraphics[width=\linewidth]{Figures/Sensitivities_AllChannels_KDE_scott_Burkert.pdf}
    \end{minipage}
    \caption{Sensitivities on the thermally-averaged dark matter self-annihilation cross-section $\langle \sigma_{\mathrm{A}} \upsilon \rangle$ as a function of the dark matter mass $m_{\mathrm{DM}}$. All annihilation channels considered for this analysis are presented for both the NFW (left) and Burkert (right) halo profiles.}
    \label{fig:Sensitivities_AllChannels}
\end{figure}

\begin{figure}[h]
    \centering
    \includegraphics[width=0.6\linewidth]{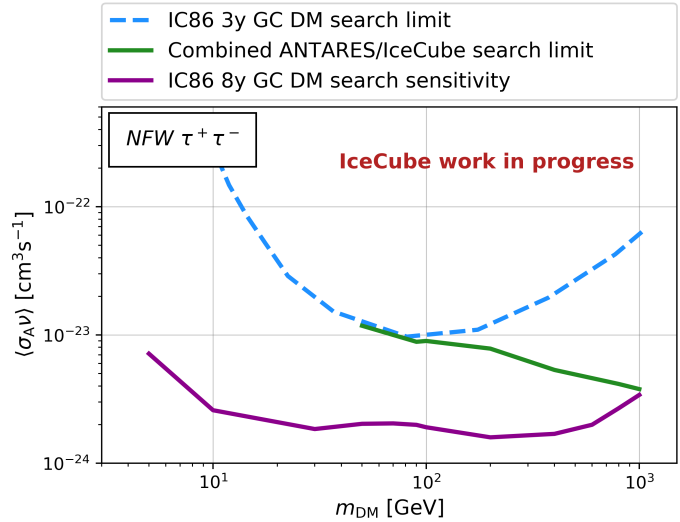}
    \caption{Sensitivity on the thermally-averaged dark matter self-annihilation cross-section $\langle \sigma_{\mathrm{A}} \upsilon \rangle$ for annihilation through $\tau^+\tau^-$ and assuming the NFW halo profile shown along with limits from IceCube~\cite{Aartsen:2017ulx} and the combined dark matter conducted with ANTARES and IceCube~\cite{Aartsen:2020tdl}}
    \label{fig:Sensitivities_Comparison}
\end{figure}

In this section, we presented the 90$\%$ CL sensitivities on the thermally-averaged dark matter self-annihilation cross-section, $\langle \sigma_{\mathrm{A}} \upsilon \rangle$, obtained for this analysis. 
These 90$\%$ CL sensitivities are shown in Figure~\ref{fig:Sensitivities_AllChannels} as a function of the dark matter mass for all evaluated annihilation channels and both considered halo profiles. The sensitivities obtained for this analysis show considerable improvement with respect to previous IceCube results~\cite{Aartsen:2017ulx}. This enhancement can be seen in Figure~\ref{fig:Sensitivities_Comparison} for the $\tau^+\tau^-$ annihilation channel and the NFW halo profile. This improvement is due to multiple factors. First, more years of data were considered for this analysis compared to the 3 years of DeepCore previously used. Furthermore, the event selection considered present a considerable improvement when compared to the previously used data set, especially at the lowest energies.
Finally, this analysis is including information about the energy and the neutrino flavour along with the angular information, which was not the case for previous IceCube analyses.

\section{Conclusion and outlooks}
\label{sec:conclusion}
We computed the sensitivities on $\langle \sigma_{\mathrm{A}} \upsilon \rangle$ for a dark matter search in the Galactic Centre using 8 years of DeepCore data. The obtained sensitivities show considerable improvements when compared to previous IceCube results from similar searches. This improvement is mainly due to the enhanced event selection considered, as well as the inclusion of the energy and the flavour information in the event distributions. As this analysis is in its final state, the final official results should soon be available. If no signal neutrinos were to be found, limits on the thermally-averaged self-annihilation cross-section $\langle \sigma_{\mathrm{A}} \upsilon \rangle$ will be computed. 

\bibliographystyle{ICRC}
\bibliography{references}



\clearpage
\section*{Full Author List: IceCube Collaboration}

\scriptsize
\noindent
R. Abbasi$^{17}$,
M. Ackermann$^{59}$,
J. Adams$^{18}$,
J. A. Aguilar$^{12}$,
M. Ahlers$^{22}$,
M. Ahrens$^{50}$,
C. Alispach$^{28}$,
A. A. Alves Jr.$^{31}$,
N. M. Amin$^{42}$,
R. An$^{14}$
% ...
\\

\noindent
% ...
$^{12}$ Universit{\'e} Libre de Bruxelles, Science Faculty CP230, B-1050 Brussels, Belgium \\
% ...
$^{14}$ Department of Physics and Laboratory for Particle Physics and Cosmology, Harvard University, Cambridge, MA 02138, USA \\
% ...
$^{17}$ Department of Physics, Loyola University Chicago, Chicago, IL 60660, USA \\
$^{18}$ Dept. of Physics and Astronomy, University of Canterbury, Private Bag 4800, Christchurch, New Zealand \\
% ...
$^{22}$ Niels Bohr Institute, University of Copenhagen, DK-2100 Copenhagen, Denmark \\
% ...
$^{28}$ D{\'e}partement de physique nucl{\'e}aire et corpusculaire, Universit{\'e} de Gen{\`e}ve, CH-1211 Gen{\`e}ve, Switzerland \\
% ...
$^{31}$ Karlsruhe Institute of Technology, Institute for Astroparticle Physics, D-76021 Karlsruhe, Germany  \\
% ...
$^{42}$ Bartol Research Institute and Dept. of Physics and Astronomy, University of Delaware, Newark, DE 19716, USA \\
% ...
$^{50}$ Oskar Klein Centre and Dept. of Physics, Stockholm University, SE-10691 Stockholm, Sweden \\
% ...
$^{59}$ DESY, D-15738 Zeuthen, Germany \\
% ...

\subsection*{Acknowledgements}

\noindent
USA {\textendash} U.S. National Science Foundation-Office of Polar Programs,
U.S. National Science Foundation-Physics Division,
U.S. National Science Foundation-EPSCoR,
Wisconsin Alumni Research Foundation,
Center for High Throughput Computing (CHTC) at the University of Wisconsin{\textendash}Madison,
Open Science Grid (OSG),
Extreme Science and Engineering Discovery Environment (XSEDE),
Frontera computing project at the Texas Advanced Computing Center,
U.S. Department of Energy-National Energy Research Scientific Computing Center,
Particle astrophysics research computing center at the University of Maryland,
Institute for Cyber-Enabled Research at Michigan State University,
and Astroparticle physics computational facility at Marquette University;
Belgium {\textendash} Funds for Scientific Research (FRS-FNRS and FWO),
FWO Odysseus and Big Science programmes,
and Belgian Federal Science Policy Office (Belspo);
Germany {\textendash} Bundesministerium f{\"u}r Bildung und Forschung (BMBF),
Deutsche Forschungsgemeinschaft (DFG),
Helmholtz Alliance for Astroparticle Physics (HAP),
Initiative and Networking Fund of the Helmholtz Association,
Deutsches Elektronen Synchrotron (DESY),
and High Performance Computing cluster of the RWTH Aachen;
Sweden {\textendash} Swedish Research Council,
Swedish Polar Research Secretariat,
Swedish National Infrastructure for Computing (SNIC),
and Knut and Alice Wallenberg Foundation;
Australia {\textendash} Australian Research Council;
Canada {\textendash} Natural Sciences and Engineering Research Council of Canada,
Calcul Qu{\'e}bec, Compute Ontario, Canada Foundation for Innovation, WestGrid, and Compute Canada;
Denmark {\textendash} Villum Fonden and Carlsberg Foundation;
New Zealand {\textendash} Marsden Fund;
Japan {\textendash} Japan Society for Promotion of Science (JSPS)
and Institute for Global Prominent Research (IGPR) of Chiba University;
Korea {\textendash} National Research Foundation of Korea (NRF);
Switzerland {\textendash} Swiss National Science Foundation (SNSF);
United Kingdom {\textendash} Department of Physics, University of Oxford.

\end{document}